\documentclass{iopjournal}
\usepackage{float}
\usepackage{amsmath}
\usepackage[numbers, sort&compress]{natbib}
\begin{document}

\articletype{Paper} 

\title{2DESR: a two-dimensional Fourier-space gyrokinetic eigenvalue code for the ion-temperature-gradient modes in tokamaks}

\author{Haochuan Wang$^1$, Jie Wang$^1$, Yuefeng Qiu, Shaojie Wang$^{*}$, Zihao Wang, Tiannan Wu, Yuesong Li, Yicheng Cai and Shiqi Xiao}

\affil{Department of Engineering and Applied Physics, University of Science and Technology of China, Hefei 230026, China}

\affil{$^1$These authors contributed equally to this work.}

\affil{$^*$Author to whom any correspondence should be addressed.}

\email{wangsj@ustc.edu.cn}


\begin{abstract}
A two-dimensional (2D) gyrokinetic eigenvalue solver, 2DESR, has been developed to solve the 2D gyrokinetic eigenvalue problem in the poloidal Fourier space for the ion-temperature-gradient (ITG) modes in tokamaks. 
With full kinetic effects of ions retained, the 2D gyrokinetic eigenvalue equations in the poloidal Fourier space have been derived and numerically solved in the 2DESR code.
In the linear ITG Cyclone test with adiabatic electrons, the 2DESR code benchmarks well against the gyrokinetic initial-value codes GENE and NLT. It is found that two branches of ITG modes coexist in the system. 
\end{abstract}

\section{Introduction} \label{sec:1}
Drift-wave (DW) instabilities have been widely investigated in plasma confinement physics, which are considered to be one of the primary mechanisms responsible for anomalous transport, leading to a degradation in confinement performance \cite{scott1990local,horton1999drift}. Among them, the ion-temperature-gradient (ITG) mode is one of the most extensively studied DW instabilities. There are two important types of codes for ITG mode simulations: initial-value codes and eigenvalue codes. Initial-value codes are powerful tools for studying nonlinear physics phenomena such as ITG turbulence \cite{villard2013global}. In the linear regime, unlike initial-value codes which require time evolution, eigenvalue codes can efficiently compute eigenvalues and eigenmode structures, which are useful for analyzing experimental observations \cite{dong2003instability,wang2024linear}. 

The eigenvalue problem for the ITG modes in tokamak plasma is essentially a two-dimensional (2D) eigenvalue problem in $(r,\theta)$ space, where $r$ and $\theta$ represent radial and poloidal coordinates, respectively. The neighboring poloidal harmonics are coupled with each other, resulting in enormous computational effort. As an asymptotic theory, the ballooning theory was developed for the 2D eigenvalue problem at high toroidal mode number $n$ \cite{connor1978shear,frieman1980general}. 
At the lowest order, the ballooning theory makes use of the translational invariance of poloidal harmonics to reduce the 2D problem to a one-dimensional (1D) problem. 
Applying the translational invariance, many 1D gyrokinetic eigenvalue solvers in ballooning-space, such as FULL \cite{rewoldt1982electromagnetic} and HD-7 \cite{dong1992toroidal}, were developed to solve for the local eigenvalue and the parallel (to the magnetic field lines) mode structure.
At the next order, the ballooning theory takes into account the mode radial envelope structures. By using the scale separation and the WKBJ approximation \cite{romanelli93radial}, the 2D eigenvalue problem is transformed into two coupled 1D eigenvalue problems (i.e., solving the parallel mode structure in the lowest order and the radial envelope in the higher order). To obtain the radial envelope, some 2D gyrokinetic eigenvalue codes \cite{lu2017symmetry,qiu2024linear} in ballooning-space were developed, in which the assumption of well-circulating particles is usually adopted to simplify the models. 

Eigenvalue codes can also be developed in the Fourier space. GLOGYSTO \cite{brunner1998global} explicitly solved the Vlasov equation by integrating analytically along the unperturbed guiding-center trajectories in a special Fourier space adapted to the toroidal geometry. This model treats passing and trapped particles separately, adopting the well-circulating assumption for the former and the deeply-trapped assumption for the latter. Recently, a five-field Landau-fluid eigenvalue code MAS \cite{bao2023mas} has been developed in the poloidal Fourier space to study multi-scale plasma problems in tokamak geometry. This model retains several important kinetic effects and neglects trapped ion and finite orbit width effects. 

Recently, a 1D gyrokinetic eigenvalue code ESR \cite{wang2024linear,wang2025development} was developed in the poloidal Fourier space by using the translational invariance, and benchmarked well against FULL and HD-7. 
However, the global structures of the ITG modes, which are related to many physics aspects of fusion interest, can not be obtained by using 1D eigenvalue solvers and can be obtained by using 2D eigenvalue solvers. 
The radial envelope of the ITG modes plays a crucial role in the mechanism of zonal flow evolution \cite{chen2000excitation,wangzi2022nonlinear}. Besides, the single-$n$ global ITG eigenmode, on the nonlinear neighboring equilibrium, such as zonal flows, restructures in the radial direction and generates multiple daughter-ballooning-modes \cite{sun_2022,sun_2023}. Moreover, to estimate the level of anomalous transport induced by ITG turbulence by using the mixing-length theory \cite{Nordman_1989,j_w_connor_2001}, the radial correlation length of the ITG turbulence can be estimated from the width of the radial envelope of the ITG modes. Therefore, it is of interest to develop a 2D gyrokinetic eigenvalue code in the poloidal Fourier space. 

In this work, we derive the 2D gyrokinetic eigenvalue equations for the ITG modes in the poloidal Fourier space with full kinetic effects for ions based on the Vlasov–Poisson system, and develop the 2DESR (2D eigenvalue solver in real space) code. In the linear ITG Cyclone test, the 2D gyrokinetic eigenvalue problem for the electrostatic ITG modes with adiabatic electrons is solved by 2DESR and the results show good agreement with those from the gyrokinetic initial-value codes GENE \cite{goerler2011global,gorler2016intercode} and NLT \cite{wang2012_1,wang2013_2,ye2016gyrokinetic,xu2017nonlinear}. 

The remaining part of this paper is organized as follows: in section \ref{sec:2}, the derivation
of the fundamental equations is presented; in section \ref{sec:3}, the numerical results are presented; in section \ref{sec:4}, the main conclusions of this work are summarized.

\section{Formulation of the basic equations} \label{sec:2}
\subsection{A brief review of the linear gyrokinetic equations}
In this work, we consider electrostatic perturbations of the ion distribution function and the electric potential in toroidal plasmas, adopting the adiabatic approximation for electrons. The linear gyrokinetic Vlasov equation for ions is as follows \cite{Brizard2007Foundations}:
\begin{equation}
\begin{aligned}
   ( \partial_t + \dot{\mathbf{X}}_0\cdot \nabla +\dot{v}_{\parallel 0}\partial_{v_\parallel}  ) \delta f_i(\mathbf{X},v_\parallel,\mu,t)=-(\dot{\mathbf{X}}_1\cdot \nabla +\dot{v}_{\parallel 1}\partial_{v_\parallel})F_{0i}.
\end{aligned}
\label{vlasov}
\end{equation}
Here, $\delta f_i$ is the perturbed distribution function for ions, $F_{0i}$ is the equilibrium distribution function for ions, $\mathbf{X}$ is the position of gyro-center, $v_\parallel$ is the parallel velocity, and $\mu$ is the magnetic moment. $\dot{\mathbf{X}}_0=v_\parallel \frac{\mathbf{B}^{*}}{B_\parallel^{*}}+\frac{\mu}{e_iB_\parallel^{*}}\mathbf{b}\times \nabla B,\dot{v}_{\parallel 0}=-\frac{\mu}{m_iB_\parallel^{*}}\mathbf{B}^{*} \cdot \nabla B,\dot{\mathbf{X}}_1=\frac{-\nabla  \langle \delta\phi  \rangle _{\mathrm{gy}} \times \mathbf{b}}{B_\parallel^{*}},\dot{v}_{\parallel 1}=-\frac{e_i}{m_i B_\parallel^{*}}\mathbf{B}^{*} \cdot \nabla \langle \delta\phi  \rangle _{\mathrm{gy}}$, with $\mathbf{B}$ the equilibrium magnetic field, $B=|\mathbf{B}|$, $\mathbf{b}=\mathbf{B}/B$, $\mathbf{B}^{*}=\mathbf{B}+\frac{m_i}{e_i}v_\parallel \nabla \times \mathbf{b},B_\parallel^{*}=\mathbf{b}\cdot \mathbf{B}^{*}$, $\delta \phi$ the perturbed electrostatic potential, $e_i$ the ion charge, and $m_i$ the ion mass. The gyro-average operator is defined as $ \langle  Q \rangle_{\mathrm{gy}} \equiv \frac{1}{2\pi}\oint Q(\xi)d\xi$, with $\xi$ the gyro-angle. 

The quasi-neutrality equation with adiabatic electrons is
\begin{equation}
\begin{aligned}
        -(\frac{e_i^2 n_{i}}{T_i}+\frac{e_e^2 n_{e}}{T_e})\delta\phi+\frac{e_i ^2}{T_i}\int \mathrm{d}^3 \mathbf{v} F_{0i} \langle\langle \delta \phi  \rangle\rangle _{\mathrm{gy}}+e_i \int \mathrm{d}^3 \mathbf{v}  \langle \delta f_i  \rangle _{\mathrm{gy}}=0,
\end{aligned}
\label{quasi}
\end{equation}
where $\int \mathrm{d}^3\mathbf{v}=\int \frac{2\pi B_\parallel^{*}}{m_i} \mathrm{d} v_\parallel \mathrm{d}\mu$, $e_e$ is the electron charge, $T_i$ and $T_e$ are equilibrium temperatures of ions and electrons, respectively, $n_i$ and $n_e$ are equilibrium densities of ions and electrons, respectively. In the next subsection, the derivations of the eigenvalue equations in the Fourier space are based on equations \eqref{vlasov} and \eqref{quasi}, with the $B_\parallel^{*}$ term \cite{hahm1988nonlinear} associated with the phase space volume conservation in gyrokinetics approximated by $B$.
\subsection{The 2D gyrokinetic eigenvalue problem in the Fourier space}
To benchmark with gyrokinetic initial-value codes, the equilibrium magnetic field employed in this work is identical to that in reference \cite{gorler2016intercode}, where the magnetic surface is concentric circular without Shafranov shift:
\begin{equation}
\begin{aligned}
    \mathbf{B}=\frac{B_0R_0}{R}(\mathbf{e}_\zeta+\frac{\epsilon_r}{q\sqrt{1-\epsilon^2_r}}\mathbf{e}_\theta),
\end{aligned}
\end{equation}
where $\zeta$ is the toroidal angle, $R_0$ is the major radius, $\epsilon_r=\frac{r}{R_0}$ is the inverse aspect ratio, $q(r)$ is the safety factor, $B_0$ is the magnetic field at the magnetic axis, and $R=R_0(1+\epsilon_r \mathrm{cos}\theta)$. We transform from usual toroidal coordinates  $(r,\theta,\zeta)$ to the flux coordinates $(r,\chi,\zeta)$, where $\chi$ is given by 
\begin{equation}
\begin{aligned}
    \chi(r,\theta)=2\mathrm{arctan}\left [\sqrt{\frac{1-\epsilon_r}{1+\epsilon_r}}\mathrm{tan}(\frac{\theta}{2})\right ],
\end{aligned}
\end{equation}
which satisfies $\frac{\mathbf{B}\cdot\nabla\zeta}{\mathbf{B}\cdot\nabla\chi}=q(r)$. 

We apply time domain Laplace transformation and Fourier transformations in the toroidal angle $\zeta$ and the poloidal angle $\chi$ to the perturbation of the physical quantity $Q(r, \chi, \zeta, t)$: 
\begin{align}
       Q(r,\chi,\zeta,t)&=\mathrm{e}^{-\mathrm{i}(n\zeta+\omega t)}\sum_m \mathrm{e}^{\mathrm{i}m\chi} \bar{Q}_{n,m}(r), \label{fourier_laplace}\\
       \bar{Q}_{n,m}(r)&=\hat{Q}_n(z,m),\label{r_z}
\end{align}
\label{fourier_transformation}
with $z \equiv nq(r)-m$. In a toroidal axisymmetric tokamak, the equations of different toroidal mode number $n$ are independent, and the subscript $n$ will be omitted in the following for simplicity. 

The gyro-average operator in the Fourier space is obtained by using the inverse Fourier transformation, which is written as
\begin{equation}
\begin{aligned}
     \langle \hat{Q} \rangle_{\mathrm{gy}} (z,m)&=\frac{1}{2\pi}\int_{-\pi}^{\pi} \mathrm{e}^{-\mathrm{i}m\chi}\mathrm{d}\chi \langle Q \rangle_{\mathrm{gy}}(r,\chi)\\
       &=\frac{1}{4\pi^2}\int_{-\pi}^{\pi} \mathrm{e}^{-\mathrm{i}m\chi}\mathrm{d}\chi \int_{-\pi}^{\pi} \mathrm{d}\xi \; Q(r+\delta r(\xi),\chi+\delta \chi (\xi))\\
        &=\frac{1}{2\pi^2} \sum_p \int_{-\pi}^{\pi}  \mathrm{d}\chi \int_0^\pi \mathrm{d}\xi  \; \mathrm{cos}[p\chi+(m+p)\delta\chi(\xi)]\hat{Q}(z+\delta z(\xi)-p,m+p).
\end{aligned}
\label{gyro_fourier}
\end{equation}
Here, $r=q^{-1}(\frac{z+m}{n})$, $\delta r=\rho \mathrm{cos}\xi$ with  $\rho= \sqrt{\frac{2\mu m_i}{e_i^2 B}}$ the Larmor radius, $\delta\chi=\chi(r+\delta r,\theta+\delta\theta)-\chi(r,\theta)$ with $\theta=2\mathrm{arctan}\left [\sqrt{\frac{1+\epsilon_r}{1-\epsilon_r}}\mathrm{tan}(\frac{\chi}{2})\right ]$ and $\delta\theta=\mathrm{arctan} \frac{\rho \mathrm{sin}\xi}{r+\rho\mathrm{cos}\xi}$, $\delta z=nq(r+\delta r)-nq(r)$.

Applying the transformation equations \eqref{fourier_laplace} and \eqref{r_z} to equation \eqref{vlasov}, with a decomposition of $\delta f_i$ in terms of its adiabatic and non-adiabatic components, $\delta f_i=-\frac{e_i F_{0i}}{T_i} \langle \delta \phi \rangle _{\mathrm{gy}}+g_i$, we obtain the Vlasov equation in the Fourier space, which is written as
\begin{equation}
\begin{aligned}
       \sum_p (\omega_p^0+\omega_p^z \partial_z+a_p \partial_{v_\parallel})\hat{g}_i(z-p,m+p,v_\parallel,\mu)=-\frac{e_i F_{Mi}^{(0)}}{T_i}\sum_p \omega_p^t \langle \delta\hat{\phi} \rangle_{\mathrm{gy}} (z-p,m+p).
\end{aligned}   
\label{vlasov_fourier}
\end{equation}
Here, the equilibrium distribution function $F_{0i}$ is taken as the local Maxwellian distribution, $F_{0i}=F_{Mi}=n_i(\frac{m_i}{2\pi T_i})^{\frac{3}{2}}\mathrm{e}^{-\frac{\frac{1}{2}m_i v_\parallel^2+\mu B}{T_{i}}}$, and $F_{Mi}^{(0)} \equiv  n_{i}\left(\frac{m_i}{2\pi T_i}\right)^{\frac{3}{2}} \mathrm{e}^{ -\frac{\frac{1}{2}m_i v_\parallel^2 + \mu B^{(0)}}{T_{i}} }$ with $B^{(0)}\equiv \frac{ B_0}{1-\epsilon_r^2}\sqrt{1+\frac{\epsilon_r^2}{q^2(1-\epsilon_r^2)}}$. The expressions for the fourier coefficients $\omega_p^0$, $\omega_p^z$, $a_p$, and $\omega_p^t$  are presented in appendix \ref{appendix_fourier_coeff}. 

The quasi-neutrality equation in the Fourier space is
\begin{equation}
\begin{aligned}
       -(\frac{e_i^2 n_{i}}{T_i}+\frac{e_e^2 n_{e}}{T_e})\delta\hat{\phi}(z,m)+e_i \frac{2\pi B^{(0)}}{m_i} \int_{-\infty}^{+\infty}\mathrm{d}v_\parallel \int_0^{+\infty}\mathrm{d}\mu\;  \bigg[ \left \langle \hat{g}_i \right \rangle _{\mathrm{gy}}(z,m,v_\parallel,\mu)& \\
       -\frac{\epsilon_r}{2}\left \langle \hat{g}_i \right \rangle _{\mathrm{gy}}(z-1,m+1,v_\parallel,\mu) -\frac{\epsilon_r}{2}\left \langle \hat{g}_i \right \rangle _{\mathrm{gy}}(z+1,m-1,v_\parallel,\mu) \bigg]=0&.
\end{aligned}
\label{quasi_fourier}
\end{equation}
Equations \eqref{vlasov_fourier} and \eqref{quasi_fourier} construct the 2D eigenvalue problem in the Fourier space. 

The boundary conditions are $\hat{g}_i(z,m,v_\parallel,\mu)|_{z,m,v_\parallel\to \pm \infty}=0$ and $\delta\hat{\phi}(z,m)|_{z,m\to \pm \infty}=0$. 
\subsection{Algebraic eigenvalue equation}\label{subsec:algebraic}
In the 2DESR code, the finite difference method is applied to solve the 2D eigenvalue problem in the Fourier space and the computational domain is $(z,m,v_\parallel,\mu)\in [-z_c,z_c]\times [m_1,m_2]\times [-v_c,v_c]\times [0,\mu_c]$, where $z_c=5,v_c=4\sqrt{\frac{2T_{\mathrm{ref}}}{m_{\mathrm{ref}}}},\mu_c=16\frac{T_{\mathrm{ref}}}{B_{\mathrm{ref}}}$ are cutoff values of $z,v_\parallel,\mu$, respectively, with $T_{\mathrm{ref}},m_{\mathrm{ref}},B_{\mathrm{ref}}$ the reference values. The coordinates $z$ and $v_\parallel$ are uniformly discretized with $\Delta z=\frac{2z_c}{N_z-1}$ and $\Delta v=\frac{2v_c}{N_{v}}$. The total number of poloidal modes $N_m=m_2-m_1+1$ is chosen based on the number of rational surfaces within the radial range of interest. The $\mu$ direction is discretized by using Gauss–Laguerre quadrature with $N_{\mu}=16$, and the nodal maximum is rescaled to $\mu_c$. In equations \eqref{gyro_fourier} and \eqref{vlasov_fourier}, we truncate at $|p|\le 3$ based on numerical convergence tests. 

After discretization, the Vlasov equation \eqref{vlasov_fourier} becomes
\begin{equation}
\begin{aligned}
       \left[\mathbf{A}(\omega)\right] [\hat{\mathbf{g}}] =\left[\mathbf{B}(\omega)\right] [\delta\hat{\boldsymbol {\phi}}],
\end{aligned}
\end{equation}
with the matrices $[\mathbf{A}(\omega)]$ and  $[\mathbf{B}(\omega)]$ given in appendix \ref{appendix_numerical_scheme}. Here, $[\hat{\mathbf{g}}]$ and $[\delta\hat{\boldsymbol {\phi}}]$ represent the discretized column vectors of the perturbed distribution function and perturbed electrostatic potential, respectively. With $N_2\equiv N_z\times N_m,N_3 \equiv N_z\times N_m\times N_v$, the column vectors $[\hat{\mathbf{g}}]$ and $[\delta\hat{\boldsymbol{\phi}}]$ are of sizes $N_3$ and $N_2$, respectively, while $[\mathbf{A}(\omega)]$ and $[\mathbf{B}(\omega)]$ are $N_3 \times N_3$ and $N_3 \times N_2$ matrices, respectively. We compute $[\mathbf{C}(\omega)] \equiv [\mathbf{A}(\omega)]^{-1} [\mathbf{B}(\omega)]$ with the sparse linear solver PARDISO, which yields
\begin{equation}
\begin{aligned}
       \left[\hat{\mathbf{g}}\right]= [\mathbf{C}(\omega)] [\delta\hat{\boldsymbol {\phi}}].
\end{aligned}
\end{equation}
Substituting $[\hat{\mathbf{g}}]$ into the quasi-neutrality equation \eqref{quasi_fourier}, we obtain the algebraic eigenvalue equation, which is written as
\begin{equation}
\begin{aligned}
       \left[\mathbf{M}(\omega)\right][\delta\hat{\boldsymbol {\phi}}]=0,
\end{aligned}
\end{equation}
with the $N_2\times N_2$ matrix $[\mathbf{M}(\omega)]$ given in appendix \ref{appendix_numerical_scheme}. The 2DESR code employs Newton's method consistent with references \cite{wang2024linear,wang2025development} to solve the algebraic eigenvalue equation for eigenvalue $\omega$. Note that the 2DESR code is parallelized along the $\mu$ direction by MPI programming and the matrix $[\mathbf{A}(\omega)]$ is stored in CSR format.
\subsection{Discussion on the choice of coordinates $(z,m,v_\parallel,\mu)$}
To capture the characteristic of each poloidal mode being radially localized on its corresponding rational surface for computational efficiency, we employ the Fourier space coordinates $(z, m)$ with $z=nq(r)-m$. 
In our previous work, the ESR code employed $(\mathcal{E},\mu)$ as its velocity coordinates \cite{wang2024linear,wang2025development}, with $\mathcal{E}$ the kinetic energy. 
The poloidal motion of the trapped ions is limited to $\theta \in [-\theta_b,\theta_b]$, with the bounce angle $\theta_b<\pi$; therefore, for the trapped ions, $g_i(|\theta|>\theta_b)=0$, and the poloidal Fourier decomposition is not convenient for the Vlasov equation. We solved the orbit-averaged Vlasov equation to obtain the perturbed density for the trapped ions there. 
In this work, the velocity coordinates are chosen to be $(v_\parallel,\mu)$, in which the distribution function $g_i$ is defined over the full poloidal range $\theta \in [-\pi,\pi]$, allowing a convenient poloidal Fourier decomposition. 

While in the $(v_\parallel,\mu)$ coordinates, the dimension of matrix $[\mathbf{A}(\omega)]$ is large (i.e., $[\mathbf{A}(\omega)]$ is a $N_3\times N_3$ matrix with $N_3 = N_z \times N_m \times N_v$), the computational effort is reasonable by using a sparse linear solver. 
For the $n=20$ ITG mode simulation with adiabatic electrons, the numbers of grid points are $(N_z,N_m,N_v,N_\mu)=(81,22,32,16)$ and the dimension of the matrix $[\mathbf{A}(\omega)]$ is $N_3=57\,024$. The simulation is performed on Intel(R) Xeon(R) Gold 6338 CPU 2.00 GHz by using $64$ cores and the total computational time is $140\ \mathrm{s}$. 

\section{Numerical results} \label{sec:3}
\subsection{Equilibrium configuration}
\begin{figure}[H]
	\centering
    \includegraphics[width=0.6\linewidth]{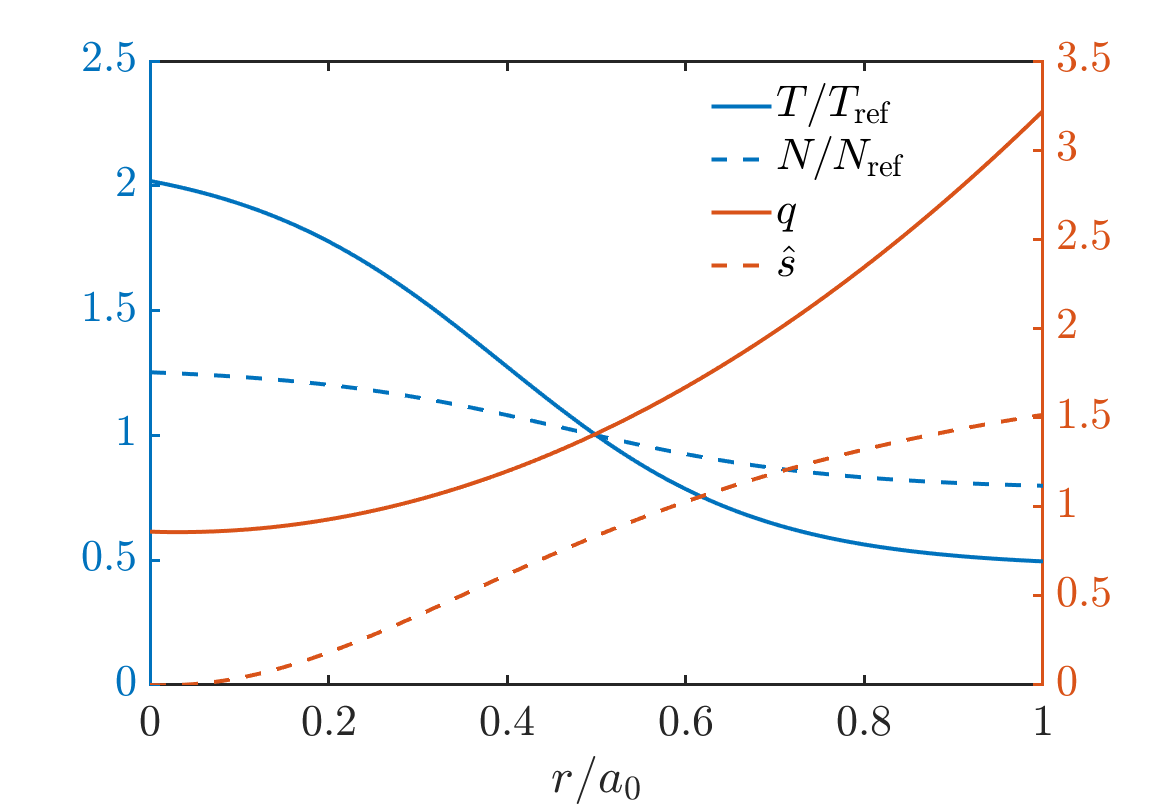}
	\caption{Equilibrium profile. Blue solid line: normalized temperature $T/T_{\mathrm{ref}}$; blue dashed line: normalized density $N/N_{\mathrm{ref}}$; red solid line: safety factor $q$; red dashed line: magnetic shear $\hat{s}=\frac{rq'(r)}{q(r)}$.}
\label{eq_profile}
\end{figure}

The equilibrium set-up in our simulations is chosen as the same Cyclone case parameters in reference \cite{gorler2016intercode}. The profile of safety factor is
\begin{equation}
\begin{aligned}
       q(r)=2.52(r/a_0)^2-0.16(r/a_0)+0.86,
\end{aligned}
\end{equation}
with $a_0$ the minor radius. The profiles of temperature and density are
\begin{align}
       T_s(r)=&T_s(r_0)\mathrm{exp}\left[ -\kappa_{Ts}\Delta_{Ts}\frac{a_0}{R_0}\mathrm{tanh}(\frac{r-r_0}{a_0\Delta_{Ts}}) \right] \label{temperature},\\
       n_{s}(r)=&n_{s}(r_0)\mathrm{exp}\left[ -\kappa_{ns}\Delta_{ns}\frac{a_0}{R_0}\mathrm{tanh}(\frac{r-r_0}{a_0\Delta_{ns}}) \right] ,\label{density}
\end{align}
where the subscript $s$ denotes electrons for $s=e$ and ions for $s=i$. Other parameters are shown in table \ref{parameter}, which reveals that the temperature and density profiles are chosen to be identical for ions and electrons. The equilibrium profiles are shown in figure \ref{eq_profile}.

\begin{table}[H]
\centering
\caption{Cyclone case parameters}\label{parameter}
\begin{tabular}{l c}
\hline
Parameter & Value \\
\hline
    $r_0/a_0$& $0.5$\\
    $a_0/R_0$& $0.60\ \mathrm{m}/1.67\ \mathrm{m}$\\ 
 $B_0$&$2.0\ \mathrm{T}$\\ 
 $T_i(r_0)/T_{\mathrm{ref}}=T_e(r_0)/T_{\mathrm{ref}}$&$1.0$\\ 
 $n_{i}(r_0)/n_{\mathrm{ref}}=n_{e}(r_0)/n_{\mathrm{ref}}$&$1.0$\\ 
 $\kappa_{Ti}=\kappa_{Te}$&$6.96$\\ 
 $\Delta_{Ti}=\Delta_{Te}$&$0.3$\\ 
 $\kappa_{ni}=\kappa_{ne}$&$2.23$\\ 
 $\Delta_{ni}=\Delta_{ne}$&$0.3$\\ 
 $m_i/m_p$&$2.0$\\ 
\hline
\end{tabular}
\end{table}
All the normalization constants are given in table \ref{reference}.
\begin{table}[H]
\centering
\caption{Reference values}\label{reference}
\begin{tabular}{l c}
\hline
   Normalization constant& Value\\
\hline
    $m_{\mathrm{ref}}/m_p$& $2.0$\\
 $e_{\mathrm{ref}}$&$1.6\times 10^{-19}\ \mathrm{C}$\\
 $B_{\mathrm{ref}}(=B_0)$&$2.0\ \mathrm{T}$\\
    $T_{\mathrm{ref}}$& $2.14\ \mathrm{keV}$\\ 
 $n_{\mathrm{ref}}$&$4.66\times 10^{19}\ \mathrm{m}^{-3}$\\ 
 $\omega_{\mathrm{ref}}(=\frac{e_{\mathrm{ref}}B_{\mathrm{ref}}}{m_{\mathrm{ref}}})$&$9.58\times10^{7} \ \mathrm{s}^{-1}$\\
 $v_{\mathrm{ref}}(=\sqrt{\frac{T_{\mathrm{ref}}}{m_{\mathrm{ref}}}})$&$3.2\times 10^5 \ \mathrm{m\ s^{-1}}$\\ 
 $L_{\mathrm{ref}}(=\frac{v_{\mathrm{ref}}}{\omega_{\mathrm{ref}}})$&$3.3\times 10^{-3}\ \mathrm{m}$\\ 
\hline
\end{tabular}
\end{table}
\subsection{2D eigenvalues and eigenmode structures of the ITG modes}
\begin{figure}[!ht]
	\centering\includegraphics[width=1\linewidth]{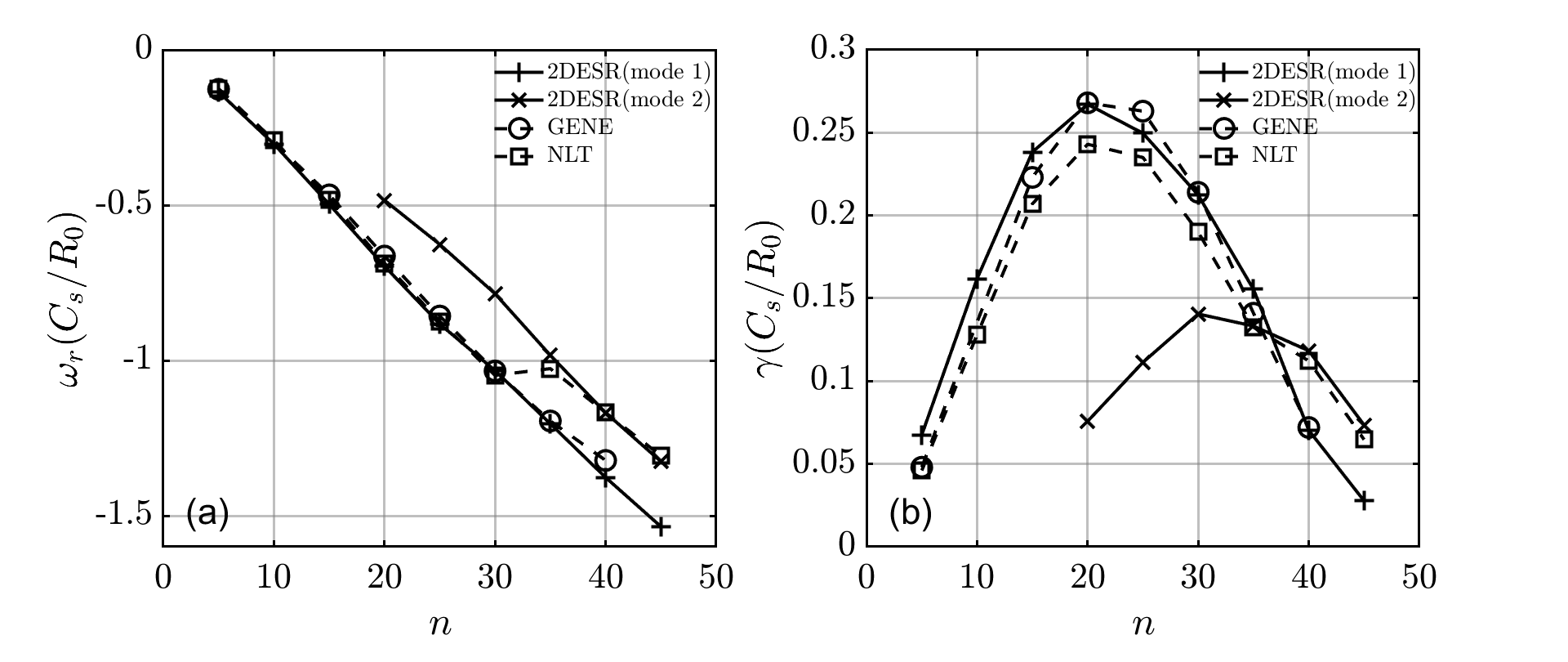}
	\caption{Real frequencies $\omega_r$ (a) and growth rates $\gamma$ (b) versus the toroidal mode number $n$, normalized by $C_s/R_0$ with $C_s=\sqrt{T_i(r_0)/m_i}$. Solid line with pluses: 2DESR(mode 1); Solid line with crosses: 2DESR(mode 2); dashed line with circles: GENE; dashed line with squares: NLT.}
    \label{scan_n}
\end{figure}

In this subsection, we scan over toroidal mode number $n=5,10,15,20,25,30,35,40,45$ to solve for the 2D eigenvalues and eigenmodes of the ITG modes with adiabatic electrons. The results are benchmarked against those from the gyrokinetic initial-value codes GENE \cite{gorler2016intercode} and NLT. 
As an eigenvalue solver, 2DESR is capable of finding multiple unstable modes. The GENE and NLT codes, as initial-value codes, always find the most unstable mode. 
Figure \ref{scan_n} shows the real frequencies and growth rates from the eigenvalue code 2DESR and those from the initial-value codes GENE and NLT. 
The 2DESR code obtains two branches of ITG modes (labeled mode 1 and mode 2), whose growth rate curves peak near $n=20$ and $n=30$, respectively. 
The results from GENE are consistent with those of the first branch, while the results from NLT are consistent with those of the most unstable branch, with both the real frequencies and the growth rates obtained by the NLT code jumping from the first branch to the second branch near $n=35$. On the other hand, a real frequency mismatch between gyrokinetic initial-value codes GENE and GKW at $n=40$ was observed, and was attributed to the possible presence of a strong sub-dominant mode \cite{gorler2016intercode}. The transition of the real frequency at $n=40$ obtained by GKW is consistent with our findings here. Since the growth rates of the two branches are close near $n=35$, resolving the most unstable branch in this region is difficult for an initial-value code. By using the NLT code, a numerically converged simulation at $n=35$ requires a total simulation time of about $2\,500\ R_0/C_s$, while the simulation at $n=20$ requires about $60\ R_0/C_s$. 

\begin{figure}[H]
	\centering
    \includegraphics[width=0.7\linewidth]{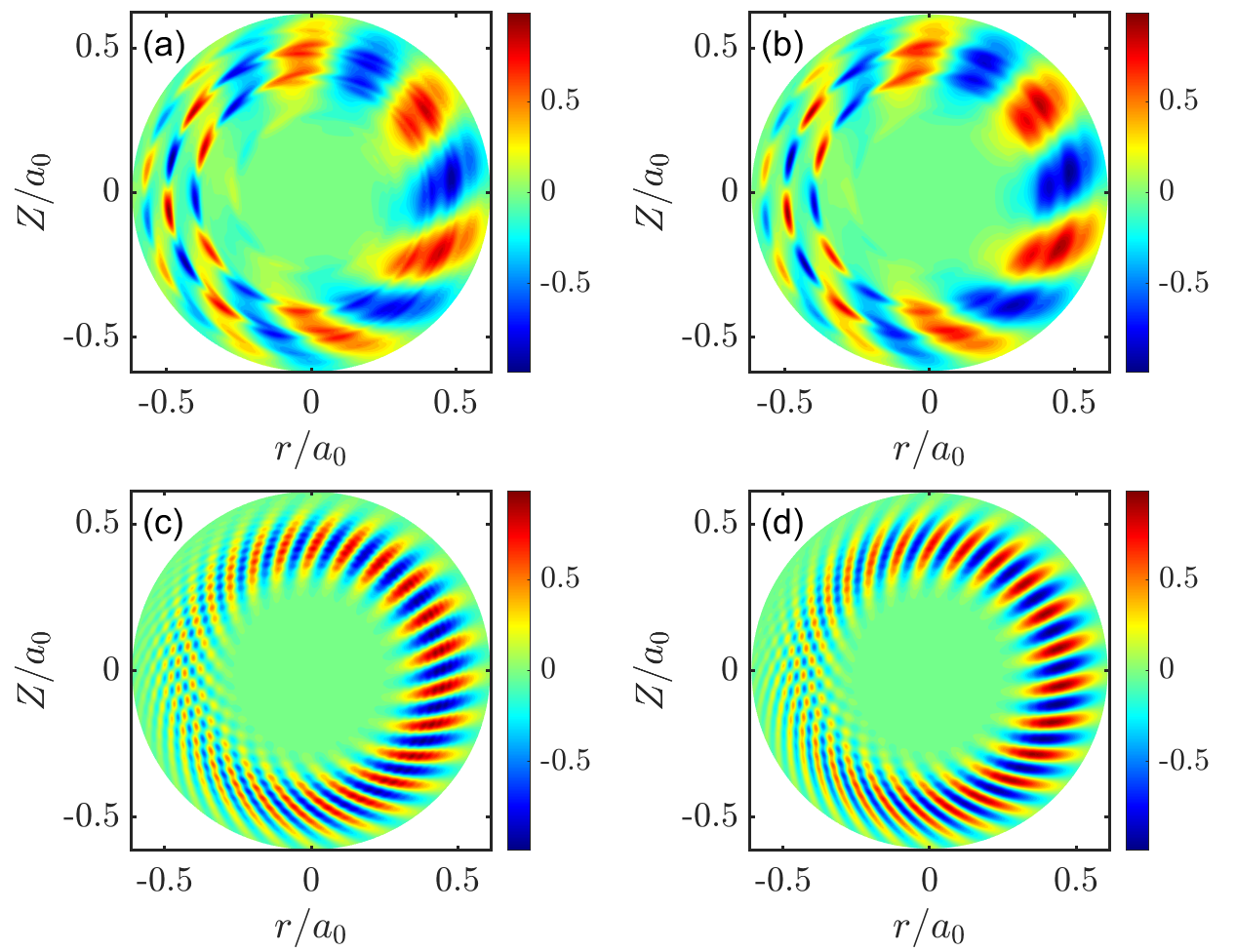}
	\caption{2D ITG eigenmode structures of electrostatic potential. (a): $n=5$, 2DESR; (b): $n=5$, NLT; (c): $n=20$, 2DESR; (d): $n=20$, NLT.}
    \label{phi2d} 
\end{figure}

\begin{figure}[H]
	\centering
    \includegraphics[width=0.8\linewidth]{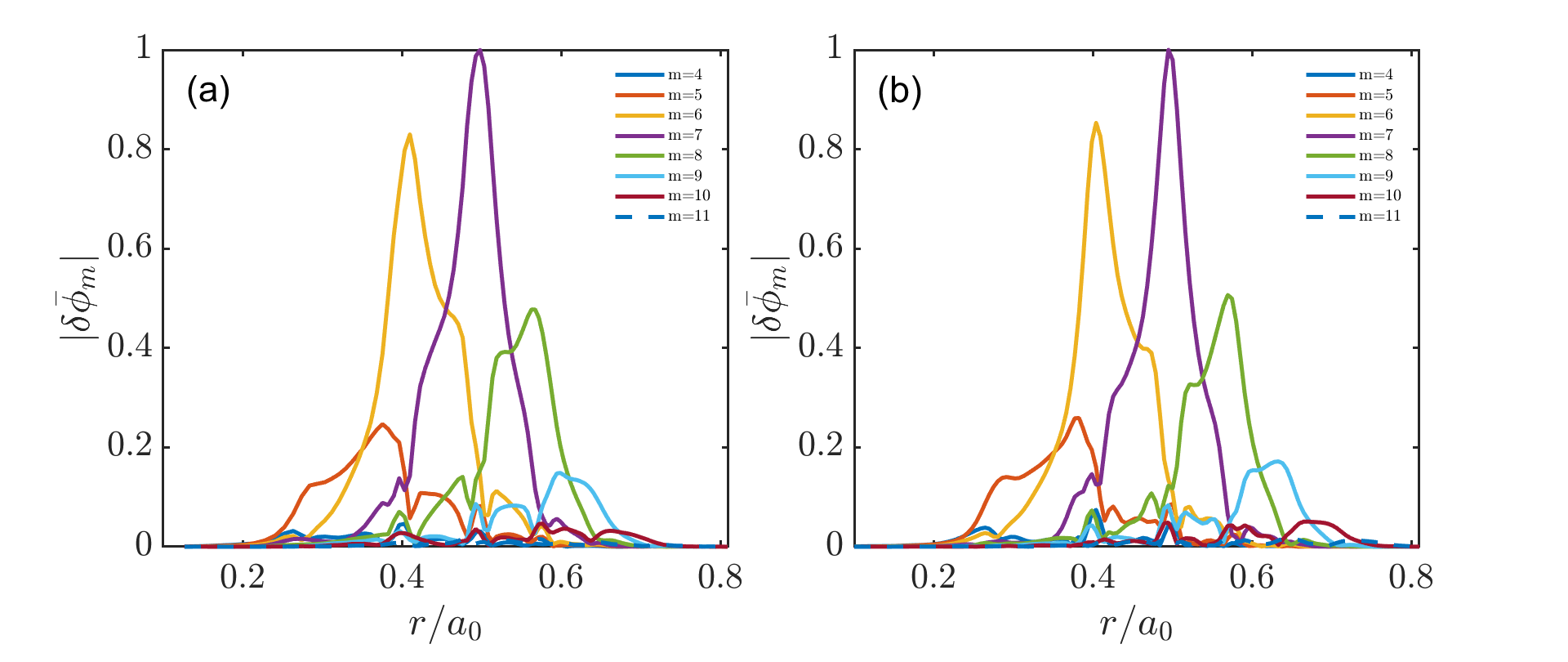}
    \includegraphics[width=0.8\linewidth]{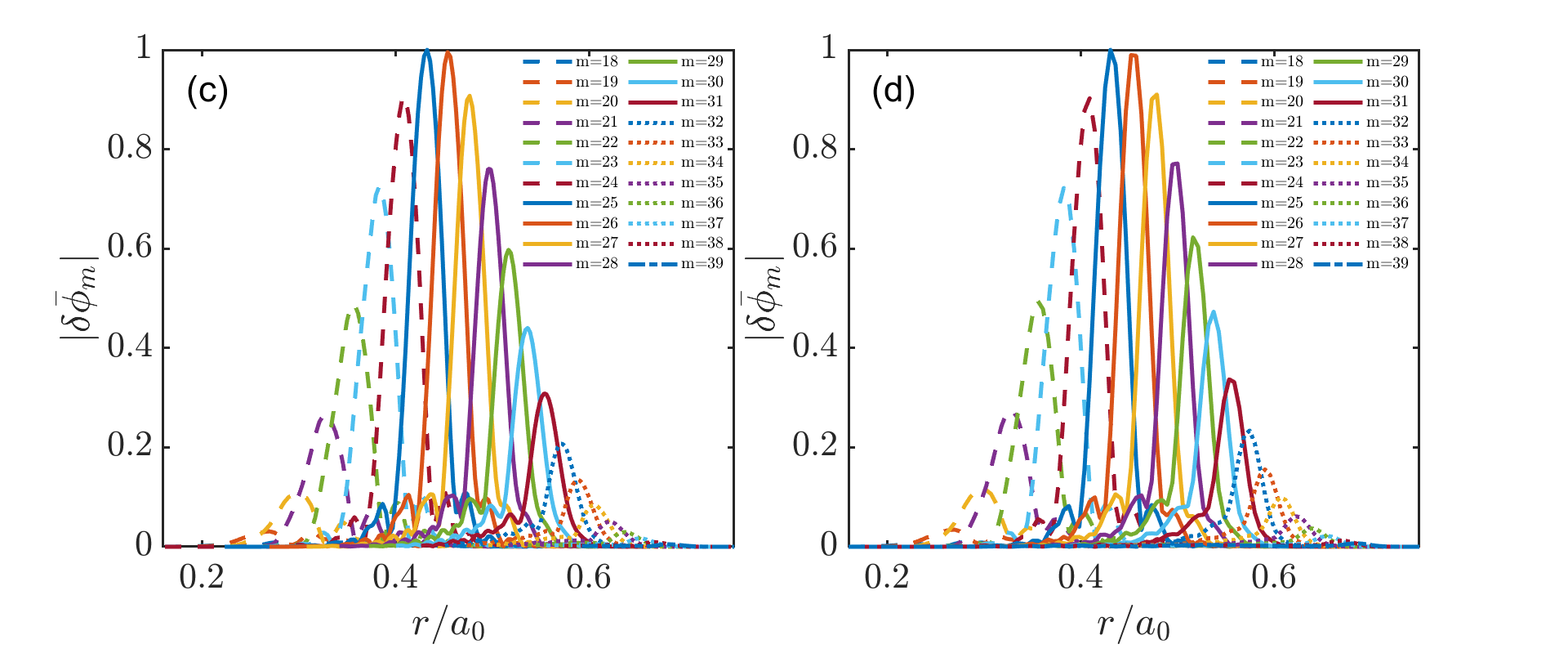}
	\caption{Radial structures of poloidal harmonics $|\delta\bar{\phi}_m|$. (a): $n=5$, 2DESR; (b): $n=5$, NLT; (c): $n=20$, 2DESR; (d): $n=20$, NLT.}
    \label{phim}
\end{figure}

We select two toroidal mode numbers, $n=5$ and $n=20$, to compare the eigenmode structures of the electrostatic potential computed by 2DESR and NLT. The 2D mode structures of the electrostatic potential are shown in figure \ref{phi2d}. The simulation results show the 2D ITG eigenmode structure with an up-down asymmetry, which is typically the tilted ballooning mode. 
The radial structures of poloidal harmonics $|\delta \bar{\phi}_m|$ are shown in figure \ref{phim}.
For $n=20$, the eigenmode is radially localized within the range $r/a_0 \in [0.3,0.6]$, where the maximum amplitude is located at $r/a_0 \approx 0.45$. 
For $n=5$, with larger distance between neighboring rational surfaces, the eigenmode is radially localized within a broader range, and exhibits apparent differences in shape between adjacent poloidal harmonics. 
For both $n=5$ (low $n$) and $n=20$ (high $n$), the eigenmode structures computed by the eigenvalue code 2DESR agree well with those computed by the initial-value code NLT. Note that the translational invariance used in the 1D eigenvalue solvers \cite{wang2024linear,rewoldt1982electromagnetic,dong1992toroidal} is not applicable at low $n$, and the 2D eigenvalue solvers \cite{lu2017symmetry,qiu2024linear} which used the second-order ballooning transformation also require high $n$ approximation, but the 2DESR code is capable of solving both low $n$ and high $n$ problems. 

\begin{figure}[H]
	\centering
    \includegraphics[width=0.9\linewidth]{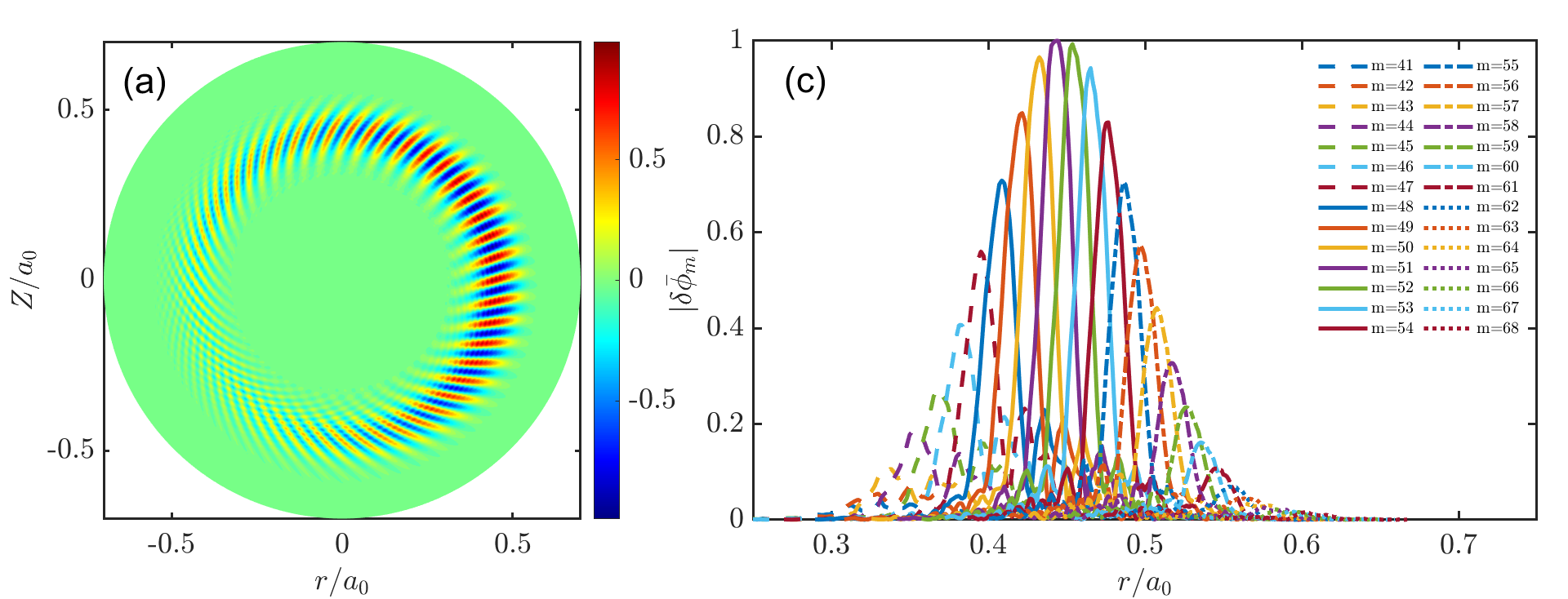}
    \includegraphics[width=0.9\linewidth]{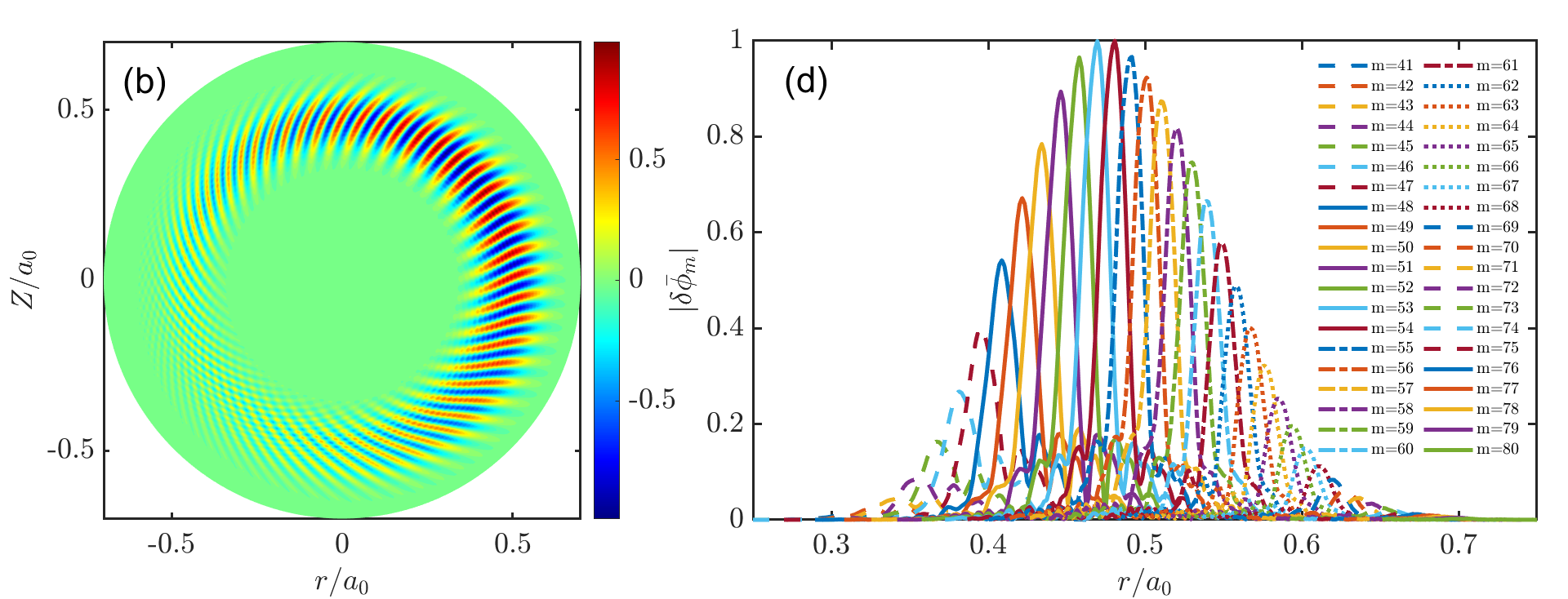}
	\caption{Eigenmode structures of electrostatic potential of two branches at $n=40$. (a)-(b): 2D mode structures of the first branch and the second branch, respectively. (c)-(d): radial structures of poloidal harmonics of the first branch and the second branch, respectively.}
    \label{phi2d_n40_2branch} 
\end{figure}

The 2D mode structures and the radial structures of poloidal harmonics of the electrostatic potential of the two branches at $n=40$ are shown in figure \ref{phi2d_n40_2branch}. The 2D structures of the two branches are both typically the tilted ballooning mode, while different radial localizations can be observed. The first branch is localized within the range $r/a_0\in [0.3,0.55]$, peaking at $r/a_0 \approx 0.45$, with approximately $28$ coupled poloidal harmonics. In contrast, the second branch is localized within a broader range of $r/a_0\in [0.3,0.65]$, peaking at $r/a_0 \approx 0.5$, with approximately $40$ coupled poloidal harmonics. 
The results from NLT agree with those of the second branch, which is the most unstable one at $n=40$. 

\section{Summary} \label{sec:4}
In this work, we have developed the 2DESR code to solve the 2D gyrokinetic eigenvalue problem for the ITG modes. The eigenvalue equations in the poloidal Fourier space with full kinetic effects of ions are derived, and the algebraic eigenvalue equation is solved numerically by using Newton's method. 

Regarding the coordinate choice, we find that the position space $(z, m)$ coordinates ($z=nq(r)-m$) help save grid points, while the velocity coordinates $(v_\parallel,\mu)$ enable a uniform treatment for passing and trapped ions in the poloidal Fourier space. The computational effort is reasonable, which is an essential prerequisite for rapid experimental analysis. 

Under the Cyclone case parameters, the eigenvalue code 2DESR benchmarks well against the gyrokinetic initial-value codes GENE and NLT in terms of both eigenvalues and eigenmode structures. By using the 2DESR code, we find that two branches of ITG modes coexist in the system, whose mode structures exhibit different radial localizations. The results from the NLT code are consistent with those of the most unstable branch. Our findings here also explain the real frequency mismatch observed in previous intercode comparison between the gyrokinetic initial-value codes GENE and GKW. 

\ack{This work was supported by National Natural Science Foundation under Grant No. 12535014, and the Strategic Priority Research Program of the Chinese Academy of Sciences under Grant Nos. XDB0790201, XDB0500302.}

\appendix
\section{Fourier coefficients in equation \eqref{vlasov_fourier}}\label{appendix_fourier_coeff}
The expressions for $\omega_p^0$ are given by
\begin{subequations}
\begin{align}
    \omega_0^0 =& \begin{aligned}[t]
    -\omega - \lambda_1^{-1}\lambda_2 \frac{v_\parallel z}{q R_0} + \lambda_1^{-\frac{1}{2}}\lambda_2^2 m \, \omega_{d0} (-\epsilon_r + \lambda_1\lambda_2^{-1}\lambda_2' R_0) + \frac{\epsilon_r}{2}\lambda_1^{-1}\lambda_2 m \frac{\mu}{e_i R_0} \\
     - \lambda_1^{-2} \lambda_2 \lambda_3 n \frac{\mu}{e_i R_0^2} (-\frac{3}{2}\epsilon_r + \lambda_1 \lambda_2^{-1}\lambda_2' R_0 ) - \lambda_2^2 n \frac{m_i v_\parallel^2}{e_i B_0 R_0} (\lambda_3 r^{-1} + \lambda_2^2 \lambda_3'), 
     \end{aligned}\\
    \omega^0_{\pm 1} =& \begin{aligned}[t]
    &\frac{\epsilon_r}{2}\lambda_1^{-1}\lambda_2 \frac{v_\parallel (z\mp 1)}{qR_0} + \frac{1}{2}\lambda_1^{-\frac{1}{2}}\lambda_2^2 (m \pm 1) \omega_{d0} - \frac{1}{2}\lambda_1^{-2}\lambda_2\lambda_3 n \frac{\mu}{e_i R_0^2} (1+\epsilon_r^2-\lambda_1\lambda_2^{-1} \lambda_2'\epsilon_r R_0) \\
     &+ \frac{\epsilon_r}{4}\lambda_1^{-1}\lambda_2^2(\lambda_4-\lambda_4^3)(m\pm 1)\frac{m_i v_\parallel^2}{e_i B_0 r R_0}, 
     \end{aligned} \\
   \omega^0_{\pm 2} =& \frac{\epsilon_r}{4}\lambda_1^{-2}\lambda_2\lambda_3 n \frac{\mu}{e_i R_0^2} + \frac{\epsilon_r}{4} \lambda_1^{-1}\lambda_2^2(-1+2\lambda_4^2-\lambda_4^4)(m\pm 2) \frac{m_i v_\parallel^2}{e_i B_0 r R_0}, \\
    \omega^0_{p} =& \frac{\epsilon_r}{4} \lambda_1^{-1}\lambda_2^2 (2\lambda_4^{|p|}-\lambda_4^{|p|-2}-\lambda_4^{|p|+2})(m+p)\frac{m_i v_\parallel^2}{e_i B_0 r R_0}, \text{ for } |p| \ge 3,
\end{align}
\end{subequations}
where 
\begin{subequations}
\begin{align}
    \lambda_1(r) &\equiv 1-\epsilon_r^2, \\
    \lambda_2(r) &\equiv \left[1 + \frac{\epsilon_r^2}{q^2\lambda_1} \right]^{-\frac{1}{2}}, \\
    \lambda_3(r) &\equiv \frac{\epsilon_r}{q\sqrt{\lambda_1}}, \\
    \lambda_4(r) &\equiv \frac{1 - \sqrt{\lambda_1}}{\epsilon_r}, \\
    \omega_{d0}(r, v_\parallel, \mu) &\equiv -\frac{\mu B^{(0)} + m_i v_\parallel^2}{e_i B_0 r R_0}.
\end{align}
\end{subequations}
The expressions for $\omega^z_p$ are given by
\begin{subequations}
\begin{align}
    \omega^z_0 &= 0, \\[6pt]
    \omega^z_{\pm 1} &= \mp \frac{nq\hat{s}}{2} \left[ \lambda_1^{-\frac{1}{2}}\lambda_2 \frac{\mu}{e_i r R_0} + (1-\lambda_4^2)\frac{m_i v_\parallel^2}{e_i B_0 r R_0} \right], \\
    \omega^z_{p} &= - \frac{p}{|p|} \cdot \frac{nq\hat{s}}{2} \lambda_2^2 \lambda_4^{|p|-1} \frac{m_i v_\parallel^2}{e_i B_0 r R_0}, \text{ for } |p| \ge 2.
\end{align}
\end{subequations}
The expressions for $a_p$ are given by
\begin{subequations}
\begin{align}
    a_{\pm 1} &= \mp \frac{1}{2}\lambda_1^{-\frac{3}{2}}\lambda_3 \frac{\mu B_0}{m_i R_0}, \\[6pt]
   a_{\pm 2} &= \pm \frac{\epsilon_r}{4}\lambda_1^{-\frac{3}{2}}\lambda_3 \frac{\mu B_0}{m_i R_0}, \\
    a_{p} &= 0, \text{ for }p = 0 \text{ and } |p| \ge 3.
\end{align}
\end{subequations}
The expression for $\omega^t_p$ is given by
\begin{equation}
\begin{aligned}
    \omega^t_{p} &= \frac{1}{2}  \left[ 2(\omega_{t0}-\omega)I_p + \omega_{t1} (I_{p-1}+I_{p+1}) \right] [\lambda_1^{\frac{1}{2}}\lambda_2^2 (m+p) + \epsilon_r \lambda_2^2 \lambda_3 n ].
\end{aligned}
\end{equation}
Here, 
\begin{subequations}
\begin{align}
    \omega_{t0}(r, v_\parallel, \mu) &\equiv \frac{T_i}{e_i B_0 r L_n} \left[ 1 + \eta_i \left( -\frac{3}{2} + \frac{\frac{1}{2}m_i v_\parallel^2 + \mu B^{(0)} }{T_i} \right) \right], \\
    \omega_{t1}(r, \mu) &\equiv -\epsilon_r \lambda_1^{-1}\lambda_2^{-1} \eta_i \frac{\mu}{e_i r L_n},
\end{align}
\end{subequations}
and $\mathrm{I}_p$ denotes the modified Bessel function of the first kind of order $p$ with argument $\epsilon_r \frac{\mu B^{(0)}}{T_i}$.
\section{Numerical solution to equations \eqref{vlasov_fourier} and \eqref{quasi_fourier}}\label{appendix_numerical_scheme}
The coordinates $z$ and $v_\parallel$ are uniformly discretized with $z_i=-z_c+(i-1)\Delta z$ $(\Delta z=\frac{1}{N})$ and $v_k=-v_c+(k-1)\Delta v$, and the partial derivatives are approximated by the central finite difference scheme, which can be written as 
\begin{equation}
\begin{aligned}
       \partial_z \hat{g}(z_i,m_j,v_{ k},\mu_l)&=\frac{1}{2\Delta z}\left[ \hat{g}(z_{i+1},m_j,v_{ k},\mu_l)-\hat{g}(z_{i-1},m_j,v_{ k},\mu_l) \right],\\
       \partial_{v_\parallel}\hat{g}(z_i,m_j,v_{ k},\mu_l)&=
       \frac{1}{2\Delta v}\left[ \hat{g}(z_{i},m_j,v_{ k+1},\mu_l)-\hat{g}(z_{i},m_j,v_{ k-1},\mu_l) \right].
\end{aligned}
\end{equation}
The velocity space integral is computed by the numerical scheme 
\begin{equation}
\begin{aligned}
       \int_{-\infty}^{+\infty}\mathrm{d}v_\parallel \int_0^{+\infty}\mathrm{d}  \mu \,\hat{g}(z_i,m_j,v_\parallel,\mu)=\sum_{k=1}^{N_{v}}\sum_{l=1}^{N_{\mu}}  \mathrm{w}_l^\mu  \Delta v \hat{g}(z_i,m_j,v_{ k},\mu_l).
\end{aligned}
\label{int_v3_numerical}
\end{equation}
Here, $\mu$ is  discretized by using Gauss–Laguerre quadrature. The grid points are $\mu_l=\frac{\mu_c x_l^L}{x^L_{N_\mu}}$ and the weights are $\mathrm{w}_l^\mu=\frac{\mu_c x_l^L \mathrm{exp}(x_l^L)}{x_{N_\mu}^L \left[ (N_\mu+1)L_{N_\mu+1}(x_l^L) \right]^2}$, with $L_{N_\mu+1}(x)$ the $(N_\mu+1)$th order Laguerre polynomial and $x_l^L$ the $l$th root of $L_{N_\mu+1}(x)$.

After discretization, the matrix $\left[\mathbf{A}^{(l)}(\omega)\right]$ for each $\mu_l$ is evaluated by
\begin{equation}
\begin{aligned}
    \left[\mathbf{A}^{(l)}(\omega)\right]_{i,j,k}^{i',j',k'} \hat{g}(z_{i'},m_{j'},v_{k'},\mu_l)=\sum_{p=-3}^{+3}  \bigg[&\omega_p^0 \hat{g}(z_{i-pN},m_{j+p},v_k,\mu_l) \\ 
    +&\omega_p^z \frac{\hat{g}(z_{i-pN+1},m_{j+p},v_k,\mu_l)-\hat{g}(z_{i-pN-1},m_{j+p},v_k,\mu_l)}{2\Delta z} \\
    + &a_p \frac{\hat{g}(z_{i-pN},m_{j+p},v_{k+1},\mu_l)-\hat{g}(z_{i-pN},m_{j+p},v_{k-1},\mu_l)}{2\Delta v} \bigg].
\end{aligned}
\end{equation}
The multi-point gyro-average is given by
\begin{equation}
\begin{aligned}
    \langle \hat{Q} \rangle_{\mathrm{gy}} (z_i,m_j)=\frac{1}{N_{\mathrm{gy}}N_{\chi}}\sum_{p=-3}^{+3}\sum_{\alpha=1}^{N_{\mathrm{gy}}}\sum_{\beta=1}^{N_{\chi}} \mathrm{cos}[p\chi_\beta+(m_j+p)\delta \chi(\xi_\alpha)]\hat{Q}(z_i+\delta z(\xi_\alpha)-p,m_j+p).
\end{aligned}
\end{equation}
Here, $\xi_\alpha=\frac{2\alpha-1}{2N_{\mathrm{gy}}}\pi$, $\chi_\beta=\frac{2\beta-1}{N_\chi}\pi$, $N_{\mathrm{gy}}=8$, and $N_{\chi}=16$. Linear interpolation is employed for sample points of $\hat{Q}(z_i+\delta z(\xi_\alpha)-p,m_j+p)$ that are not on $z$-grid points. Consequently, the gyro-average matrix $\left[ \mathbf{J}_0^{(l)} \right]$ for each $\mu_l$ can be obtained from
\begin{equation}
\begin{aligned}
    \langle \hat{Q} \rangle_{\mathrm{gy}} (z_i,m_j)=\left[ \mathbf{J}_0^{(l)} \right]_{i,j}^{i',j'} \hat{Q}(z_{i'},m_{j'}).
\end{aligned}
\end{equation}
Then, the matrix $\left[\mathbf{B}^{(l)}(\omega)\right]$ for each $\mu_l$ is evaluated by 
\begin{equation}
\begin{aligned}
    \left[\mathbf{B}^{(l)}(\omega)\right]_{i,j,k}^{i',j'}\delta \hat{\phi}(z_{i'},m_{j'})=-\frac{e_i F_{Mi}^{(0)}}{T_i}\sum_{p=-3}^{+3}\omega_p^t \left[\mathbf{J}_0^{(l)}\right]_{i-pN,j+p}^{i',j'} \delta\hat{\phi} (z_{i'},m_{j'}).
\end{aligned}
\end{equation}
After obtaining $\left[\mathbf{C}^{(l)}(\omega)\right]=[\mathbf{A}^{(l)}(\omega)]^{-1} [\mathbf{B}^{(l)}(\omega)]$ with the sparse linear solver PARDISO, the matrix $\left[\mathbf{M}(\omega)\right]$ is evaluated by 
\begin{equation}
\begin{aligned}
    \left[\mathbf{M}(\omega)\right]_{i,j}^{i',j'}\delta \hat{\phi}(z_{i'},m_{j'})=&-(\frac{e_i^2n_i}{T_i}+\frac{e_e^2n_e}{T_e})\delta \hat{\phi}(z_i,m_j)
    +e_i\frac{2\pi B^{(0)}}{m_i}\sum_{k=1}^{N_v}\sum_{l=1}^{N_\mu}\mathrm{w}_l^\mu \Delta v \bigg\{ \left[\mathbf{J}_0^{(l)}\right]_{i,j}^{i',j'}\\&-
    \frac{\epsilon_r}{2}\left[\mathbf{J}_0^{(l)}\right]_{i-N,j+1}^{i',j'}-\frac{\epsilon_r}{2}\left[\mathbf{J}_0^{(l)}\right]_{i+N,j-1}^{i',j'}\bigg\}\left[\mathbf{C}^{(l)}(\omega)\right]_{i',j',k}^{i'',j''}\delta \hat{\phi}(z_{i''},m_{j''}).
\end{aligned}
\end{equation}

\endappendix
\bibliography{iopjournal}

\end{document}